# Atmospheric-Pressure Ar/Air Plasma Jet-Induced Degradation of Azo Dyes in Aqueous Solutions: Kinetic and Mechanistic Insights

Mohammed Shihab[1,2], Alaa El-Ashry[1], Seham A. Ibrahim[3], Sarah Salah[4], Abdelhamid Elshaer[4], Nabil El-Siragy[1], Atef A. Elbendary[1]

[1] Department of Physics, Faculty of Science, Tanta University, Tanta 31527, Egypt

[2] Department of Physics, Basic Science, Alsalam University, Gharbia 31711, Egypt

[3] Chemistry Department, Faculty of Science, Tanta University, Tanta, 31527, Egypt

[4] Department of Physics, Faculty of Science, Kafrelsheik University, Kafr-Elsheik, Egypt

E-mail: Mohammed.Shihab@science.tanta.edu.eg

## Abstract

Atmospheric-pressure non-thermal plasmas are promising platforms for advanced oxidation in water treatment, yet quantitative coupling between reactive species delivery, solution chemistry, and molecular fragmentation remains unclear. We investigate degradation of two structurally related azo dyes using an Ar plasma jet in a plasma–liquid discharge configuration with an immersed counter-electrode to enhance interfacial coupling. Plasma exposure generated a reactive oxygen and nitrogen species environment and strong acidification, increasing proton concentration up to 49-fold. UV–Vis analysis showed rapid chromophore decay, achieving 88% and 94% removal within 40 min. Biphasic kinetics indicated a transition from radical-flux-controlled to transport-influenced regimes. Fluorescence and Raman spectroscopy confirmed transient oxidized intermediates and progressive π-conjugation breakdown, elucidating plasma-driven oxidative fragmentation mechanisms.

## 1- Introduction

The release of azo dyes from the textile, printing, and pharmaceutical sectors poses a significant environmental issue due to their intricate aromatic structures, high stability, and the potential to form toxic intermediates during incomplete degradation. Traditional physicochemical and biological treatments often demonstrate limited mineralization efficiency and may cause secondary pollution, thereby prompting the investigation of advanced oxidation processes (AOPs) that can generate highly reactive species for effective pollutant degradation. Among the emerging AOPs, non-thermal atmospheric-pressure plasma (APP) technologies have shown considerable potential for water treatment, as they can produce reactive oxygen and nitrogen species (RONS) in situ—such as hydroxyl radicals, ozone, and hydrogen peroxide—without the need for external chemical additives [1,2]. Recent research has indicated that plasma-based advanced oxidation processes can achieve rapid degradation of dye and antibiotic wastewater through controlled RONS production and enhanced interfacial mass transfer [3,4]. Atmospheric-pressure plasma jets have demonstrated a strong dependence of degradation efficiency on operational parameters and plasma–liquid coupling [5]. Furthermore, comparative studies of dye degradation kinetics under pulsed plasma conditions have revealed mechanistic transitions between radical-dominated and diffusion-controlled regimes [6].

Plasma-activated water (PAW) and plasma jets have been demonstrated as efficient oxidative media for degrading diverse organic pollutants including dyes, pharmaceuticals, and persistent contaminants, with tunable reactive species profiles achieved through reactor design and operating parameters [1,4]. Lamichhane et al. recently reported enhanced dye degradation using plasma jet-generated bubbles, where improved •OH diffusion significantly increased removal efficiency and structural transformation of dye molecules, as confirmed by spectroscopic analysis [7].

Despite these advances, most studies predominantly focus on overall decolorization or bulk removal metrics, offering limited mechanistic insight into intermediate formation or molecular fragmentation pathways. Furthermore, although APPJs and DBD systems have been extensively explored, there remains a need for integrated spectroscopic diagnostics to correlate plasma-induced solution acidification, reactive species dynamics, and the structural breakdown of complex dye chromophores during treatment [1,8,9,10]. Reactor geometry and plasma–liquid interaction mode are crucial in determining species transport, interfacial coupling, and mass transfer efficiency. Plasma jets with submerged electrodes or bubble-enhanced conditions provide pathways to intensify charge transfer and reactive species penetration into the liquid bulk. However, these approaches have been less explored in the context of the stepwise oxidative fragmentation of structurally complex azo dyes [7,11,12].

In this study, we systematically explore the degradation of two structurally related azo dyes, MS16 and MS17, as depicted in Figure 1, utilizing an atmospheric-pressure Ar plasma jet configured in a plasma–liquid discharge mode. The novelty of this research lies in: (i) employing a plasma jet with an immersed counter-electrode to enhance the delivery of reactive species and electric field coupling; (ii) quantitatively correlating plasma-induced acidification with degradation kinetics; (iii) monitoring transient fluorescence enhancement and spectral shifts as indicators of intermediate oxidation states; and (iv) providing Raman spectroscopic evidence of aromatic conjugation breakdown and oxygen functionalization. By comparing the degradation behavior of these two closely related molecular structures, we elucidate how substituent chemistry influences plasma-driven oxidation pathways, thereby advancing the fundamental understanding necessary to optimize plasma systems for scalable wastewater remediation.

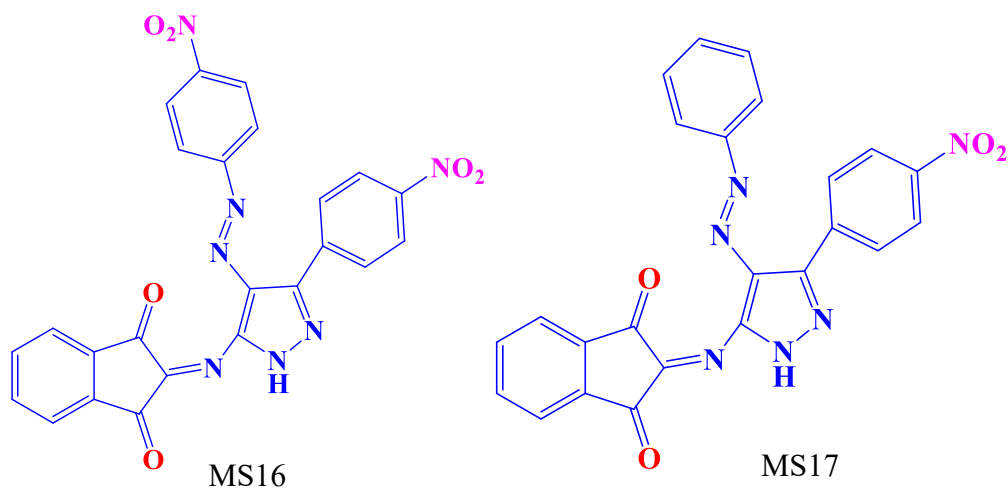

**Figure 1**. MS16 is (*E*)-2-((3-(4-nitrophenyl)-4-((4-nitrophenyl)diazenyl)-1H-pyrazol-5-yl)imino)-1*H*-indene-1,3(2*H*)-dione and MS17 is (*E*)-2-((3-(4-nitrophenyl)-4-(phenyldiazenyl)-1H-pyrazol-5-yl)imino)-1*H*-indene-1,3(2*H*)-dione.

## 2- Atmospheric-Pressure Plasma Jet (APPJ) System

Figure 2 illustrates the experimental configuration of an atmospheric-pressure plasma jet (APPJ) system employed for the degradation of organic dyes in liquid media. High voltage (HV, 4 kV, pulsed 30 kHz) is supplied to the powered electrode integrated within the plasma jet assembly, while a counter electrode is immersed in the dye solution, forming a plasma–liquid discharge configuration. Argon (Ar) is used as the working gas at a controlled flow rate of 1 L·min$^{-1}$ and is injected through the jet nozzle toward the liquid surface. Upon application of high voltage, a non-thermal plasma plume is generated at atmospheric pressure and propagates along the Ar flow direction. The plasma plume impinges on the liquid surface, establishing a localized plasma–liquid interaction zone. The electric field extends into the liquid bulk via the submerged electrode, enhancing charge transfer and reactive species generation within the solution. The dye solution is prepared by dissolving the organic dye in alcohol under mild heating to ensure complete solubilization. The concentrated solution is subsequently diluted with deionized water to obtain the desired working concentration prior to plasma treatment. Within the liquid phase, the plasma–liquid interaction produces a variety of reactive oxygen and nitrogen species (RONS), including hydroxyl radicals (•OH), hydrogen peroxide ($H_2O_2$), superoxide ($O_2$•$^-$), and other oxidative intermediates. These species diffuse into the bulk solution, initiating oxidative degradation of the dye molecules. The schematic arrows in the liquid represent convective flow and mixing induced by gas injection and localized heating at the plasma–liquid interface, which promote mass transport and enhance degradation kinetics. Overall, the system operates as a plasma-driven advanced oxidation process (AOP), where reactive species generated in the gas phase and at the interface play a dominant role in dye decomposition.

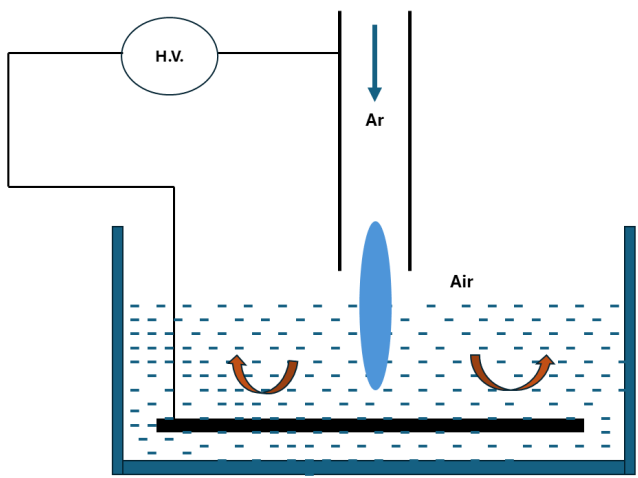

**Figure 2.** Schematic Description of the Atmospheric-Pressure Plasma Jet Reactor for Dye Degradation

**3- Plasma-Induced Acidification During Atmospheric-Pressure Treatment**

The evolution of solution pH during atmospheric-pressure Ar plasma jet treatment reveals pronounced acidification for both dye systems (MS16 and MS17). As shown in Figure 3, MS16 exhibits a gradual decrease in pH from 6.34 at t = 0 min to a minimum of 4.65 at 30 min, followed by slight stabilization at 4.70 at 40 min. In contrast, MS17 shows a sharper initial decline, decreasing from 6.18 to 5.02 within the first 20 min, reaching a minimum of 4.75 at 30 min, and subsequently increasing slightly to 4.90 at 40 min. To quantify the extent of acidification, pH values were converted to proton concentrations. For MS16, [$H^+$] increased from $4.57 \times 10^{-7}$ M to $2.24 \times 10^{-5}$ M by 30 min, corresponding to an approximately 49-fold increase. For MS17, [$H^+$] increased from $6.61 \times 10^{-7}$ M to $1.78 \times 10^{-5}$ M, corresponding to an approximately 27-fold increase. Although MS17 acidifies more rapidly during the initial treatment period, MS16 ultimately exhibits the larger net proton accumulation. The observed acidification is characteristic of plasma–liquid interactions under ambient air conditions. Energetic electrons and excited argon species (Ar*) generated in the plasma initiate dissociation and excitation processes involving $O_2$, $N_2$, and $H_2O$ near the gas–liquid interface. These reactions produce reactive oxygen and nitrogen species (RONS), including O, $O_2^{•-}$, •OH, NO, $NO_2$, $O_3$, and secondary species such as $H_2O_2$, $HNO_2$, and $HNO_3$.

Dissolution and hydration of $NO_x$ species in the liquid phase lead to the formation of nitrous and nitric acids, while recombination of hydroxyl radicals yields hydrogen peroxide. The cumulative formation of these acidic species results in progressive proton release and solution acidification.

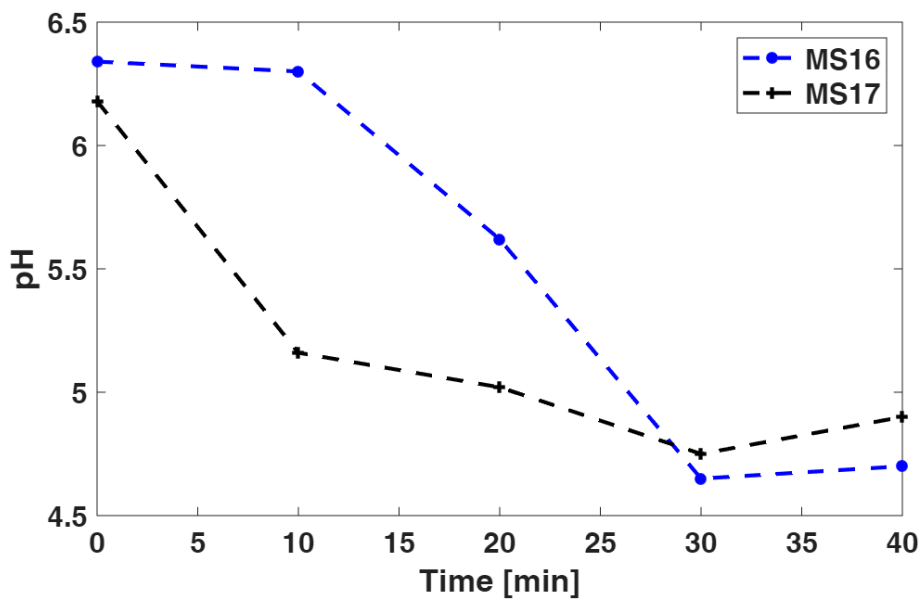

**Figure 3.** Temporal evolution of pH during atmospheric-pressure Ar plasma jet treatment of MS16 and MS17 dye solutions.

The temporal behavior of pH suggests two kinetic regimes. During the first 10–20 min, rapid RONS production at the interface dominates, leading to a steep decline in pH. This stage is governed by plasma-driven radical generation and mass transfer into the liquid phase. After approximately 30 min, the system approaches a quasi-steady state in which the rate of RONS formation is balanced by recombination, diffusion losses, and chemical consumption within the solution. The slight increase in pH observed after 30 min for both dyes may reflect partial neutralization, buffering by dye molecules or degradation intermediates, or equilibrium effects associated with accumulated species such as $NO_2^-$, $NO_3^-$, and $H_2O_2$.

The distinct acidification profiles of MS16 and MS17 indicate that molecular structure influences plasma–liquid chemistry. The faster initial pH drop observed for MS17 suggests higher reactivity toward incoming radicals or reduced initial buffering capacity. Conversely, the greater overall proton accumulation in MS16 implies sustained acid-forming reactions or differences in secondary oxidation pathways. Variations in functional groups, ionizable substituents, and aggregation behavior likely modulate radical scavenging efficiency and interfacial reaction kinetics. These results demonstrate that plasma-induced acidification is not merely a secondary effect but an integral component of the degradation environment. The substantial increase in proton concentration (up to nearly two orders of magnitude) enhances oxidative and hydrolytic processes, potentially accelerating chromophore cleavage and structural modification of the dye molecules. Overall, atmospheric-pressure Ar plasma treatment generates significant acidification through RONS formation at the gas–liquid interface, and the magnitude and kinetics of this process depend strongly on dye-specific chemical properties.

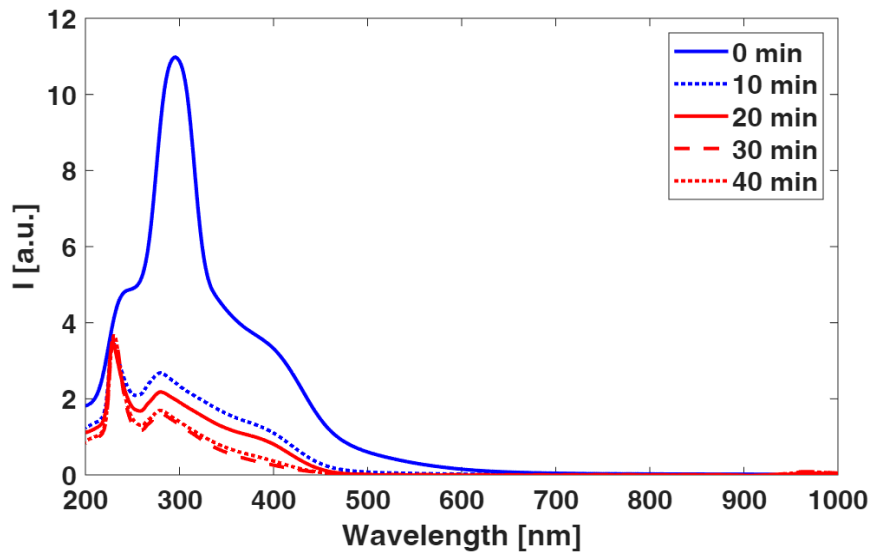

**Figure 4.** Time-resolved UV-Vis absorption spectra highlighting the degradation of MS16.

**4- Plasma-assisted oxidative degradation and Kinetic Analysis**

The degradation behavior of MS16 and MS17 under atmospheric-pressure Ar plasma jet treatment was evaluated by monitoring the temporal evolution of their UV–Vis absorption spectra. As shown in Figures 4 and 5, a pronounced and continuous decrease in the characteristic absorption bands is observed with increasing treatment time, clearly indicating progressive destruction of the chromophoric system. Both dyes display three principal absorption regions: (i) a strong band near ~290 nm attributed to π→π* transitions within the aromatic rings and extended conjugated framework; (ii) a higher-energy band around ~230 nm associated with π→π* transitions of substituted phenyl and heteroaromatic moieties; and (iii) a visible-region band centered near ~410 nm for MS16 and ~454 nm for MS17, corresponding to the azo (–N=N–) chromophore responsible for the dye coloration.

Upon plasma exposure, all absorption bands decrease steadily in intensity without noticeable bathochromic shifts or the formation of new stable absorption features in the visible region. The rapid attenuation of the visible azo band provides direct evidence of –N=N– bond cleavage and consequent collapse of the π-conjugated system. Because the intense color of azo dyes arises from electron delocalization across the aromatic–azo–aromatic framework, disruption of the azo linkage leads to immediate decolorization. Simultaneously, the progressive reduction of the UV bands (~290 nm and ~230 nm) indicates that degradation extends beyond simple chromophore cleavage to include oxidative modification of the aromatic rings and heterocyclic (pyrazole) units. The absence of persistent secondary absorption peaks suggests that stable aromatic intermediates do not accumulate in significant amounts, supporting a mechanism of extensive oxidative fragmentation rather than partial structural rearrangement. This spectral evolution is consistent with plasma-driven advanced oxidation processes, in which high-energy electrons, excited argon species (Ar*), and interfacially generated reactive oxygen and nitrogen species (RONS)—including •OH, O, $O_2^{•-}$, $O_3$, $H_2O_2$, and NOx-derived species—collectively attack the dye molecules. Among these oxidants, hydroxyl radicals (E° ≈ 2.8 V) exhibit near diffusion-controlled reactivity toward aromatic compounds, promoting hydroxylation, electron abstraction, and azo bond scission. These reactions progressively shorten the effective conjugation length and dismantle the chromophoric framework.

Overall, the combined attenuation of visible and ultraviolet absorption bands confirms that plasma treatment induces stepwise oxidative degradation, beginning with azo bond cleavage and followed by hydroxylation and fragmentation of aromatic and pyrazole moieties. The lack of long-lived chromophoric intermediates indicates that the process advances toward smaller, weakly absorbing or non-absorbing fragments, consistent with deep oxidative transformation rather than superficial bleaching.

## 4.1. Spectroscopic Evolution

The intensity of the primary absorption band at 290 nm for both samples decreased significantly over the 40-minute treatment period. Sample MS16 showed a progressive reduction, reaching a residual absorbance of approximately 12% of its initial value ($C/C_0$ = 0.12) by the end of the reaction. MS17 demonstrated superior catalytic kinetics, with a sharp drop in absorbance occurring within the first 5 minutes. The almost complete disappearance of the characteristic UV signature by time of 40 min suggests the effective cleavage of the chromophore groups and potential mineralization of the organic pollutants into colorless intermediates. UV–Vis spectroscopy remains a primary tool for monitoring chromophore cleavage in plasma-driven dye degradation systems. Similar rapid attenuation of characteristic azo absorption bands has been reported in plasma and UV-assisted advanced oxidation studies, where the decrease in $\lambda_{max}$ intensity correlates with aromatic ring rupture and π-conjugation breakdown [11,12]. The absence of emergent stable absorption bands in the visible region suggests progressive mineralization rather than formation of persistent aromatic intermediates, consistent with findings in non-thermal plasma oxidation systems [6].

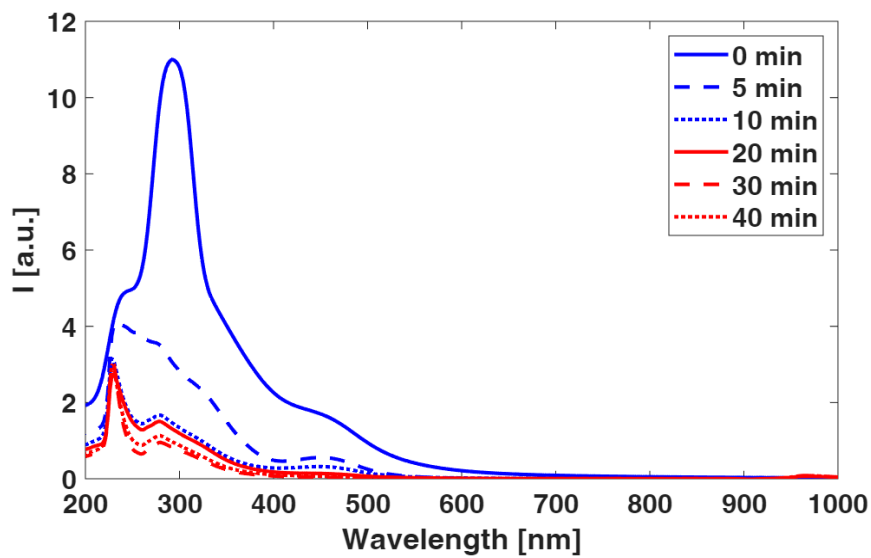

**Figure 5.** Time-resolved UV-Vis absorption spectra highlighting the degradation of MS17.

## 4.2. Comparative Degradation Kinetics

To further quantify the degradation efficiency, the relative concentration ($C/C_0$) was plotted as a function of time (Figure 6 and table 1), where $C_0$ is the initial concentration ($C_0=0.5\times10^{-3}$ M) and C is the concentration at different times, determined via the Beer-Lambert Law [13-17].

Table (1). Efficiency of removing the dyes.

| Parameter | MS16 | MS17 |
|---|---|---|
| Initial $C/C_0$ (0 min) | 1.00 | 1.00 |
| $C/C_0$ at 10 min | 0.29 | 0.17 |
| Final $C/C_0$ (40 min) | 0.12 | 0.06 |
| Total Efficiency (%) | 88.0% | 94.0% |

As illustrated in Figure 6, both samples exhibit a biphasic degradation profile: a rapid initial decay followed by a slower asymptotic phase. The enhanced performance of MS17 (94% degradation) compared to MS16 (88% degradation) can be attributed to the higher reaction rate during the first 10 minutes. In the case of MS17, the concentration plummeted to $C/C_0 = 0.17$ within 10 minutes, whereas MS16 required nearly 20 minutes to reach a similar degradation level.

### 4.3. Reaction Mechanism Insights

The accelerated degradation rate observed in MS17 suggests a more efficient generation of reactive oxygen species (ROS), such as hydroxyl radicals •OH or superoxide anions ($O_2$•⁻). The absence of significant new absorption bands in the visible or UV range during the process indicates that the dye molecules are likely undergoing a thorough breakdown rather than transforming into stable secondary aromatic by-products.

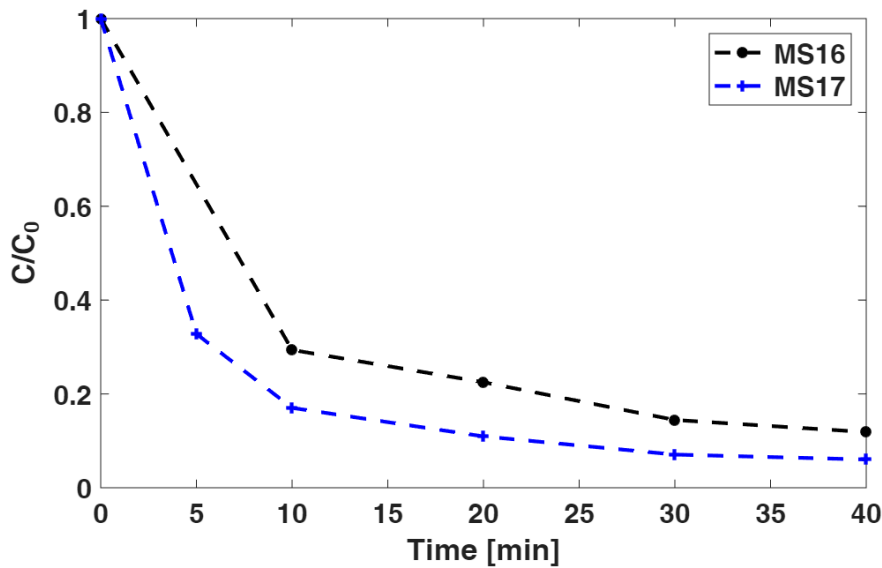

**Figure 6.** The temporal evolution of the normalized concentration ($C/C_0$) determined by the Beer-Lambert Law.

### 5. Fluorescence Evolution During Plasma Treatment

The fluorescence spectra of MS16 and MS17 in figure 7 exhibit pronounced temporal evolution during atmospheric-pressure Ar plasma exposure. In both systems, the untreated samples (0 min) show the lowest emission intensity across the spectral range, while plasma-treated samples display progressively enhanced fluorescence, reaching a maximum at approximately 30 min, followed by slight stabilization or a minor decrease at 40 min. For MS16, fluorescence intensity increases sharply from 10 to 30 min, reaching a maximum (~6.7 × 10³ a.u.), then decreases moderately at 40 min. MS17 follows a similar but less pronounced trend, with a maximum intensity of ~4.6 × 10³ a.u. at 30 min and near stabilization thereafter. Overall, MS16 exhibits significantly higher emission intensity than MS17 under comparable treatment conditions. In addition to intensity changes, both dyes display a measurable blue shift in emission maxima from ~486 nm (untreated) toward ~425–430 nm after plasma exposure. The shift indicates chemical modification of the chromophore rather than simple concentration effects. A hypsochromic shift is consistent with shortening of the effective π-conjugation length or formation of oxidized fragments with altered electronic structure. The increase in fluorescence up to 30 min is noteworthy, as simple degradation of the parent dye would typically reduce emission intensity. Instead, the observed enhancement suggests that plasma treatment induces oxidative transformation pathways leading to the formation of new fluorescent species with higher quantum yield. Reactive oxygen and nitrogen species (RONS) generated by the Ar plasma jet in ambient air (e.g., •OH, O, NOx, $H_2O_2$) likely attack the dye molecules, producing partially oxidized or fragmented products that retain or enhance emissive properties. Fluorescence spectroscopy has been increasingly employed to track oxidative transformation of organic pollutants in aqueous systems due to its sensitivity to subtle structural modifications [18]. Plasma-induced oxidative processes can generate intermediate species with altered electronic structures and enhanced quantum yields prior to complete fragmentation. Similar fluorescence enhancement followed by quenching has been observed during advanced oxidation of organic contaminants, reflecting sequential formation and destruction of emissive intermediates [10].

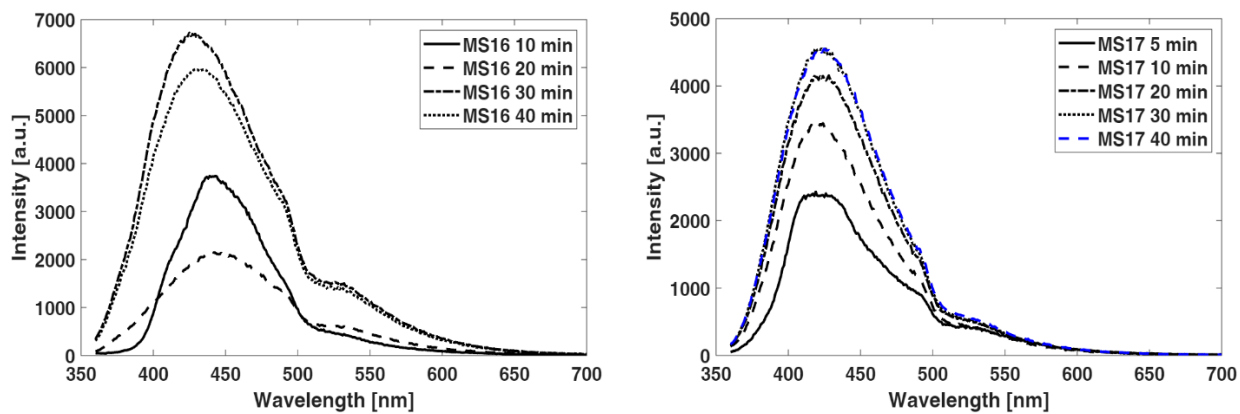

**Figure 7.** Fluorescence spectra of MS16 (left) and MS17 (right) recorded at different atmospheric-pressure Ar plasma treatment times.

The concurrent plasma-induced acidification of the solution further supports this interpretation. The decrease in pH promotes protonation of functional groups and modifies electronic distribution within the dye molecules or their degradation intermediates. Protonated species often exhibit altered radiative/non-radiative decay rates, which may contribute to the enhanced fluorescence intensity observed during intermediate treatment times. The slight intensity decrease at 40 min suggests the onset of over-oxidation or further fragmentation of the initially formed fluorescent intermediates into smaller, weakly emissive or non-emissive species. Thus, the fluorescence evolution reflects two competing processes: (i) formation of highly emissive oxidized intermediates and (ii) subsequent degradation of these intermediates at prolonged exposure.Comparatively, MS16 exhibits stronger fluorescence enhancement than MS17, indicating greater susceptibility to plasma-induced formation of emissive products or differences in molecular structure, conjugation stability, or solvent interaction. Despite quantitative differences, both dyes follow the same mechanistic trend: plasma exposure initially promotes formation of fluorescent transformation products, followed by gradual over-degradation at extended treatment times. Overall, the combined fluorescence, UV–Vis degradation, and pH results demonstrate that atmospheric-pressure plasma does not simply quench the dye emission through direct destruction. Rather, it induces stepwise chemical transformation, generating transient fluorescent species before ultimate breakdown. This behavior highlights the complex interplay between oxidative fragmentation, protonation effects, and plasma–liquid interfacial chemistry during dye treatment.

## 6. Raman spectroscopy of Precipitations

Raman spectroscopy provides molecular-level insight into structural reorganization during plasma oxidation. Real-time Raman monitoring of plasma-treated aqueous systems has demonstrated its capability to resolve aromatic bond cleavage, carbonyl formation, and oxygen functionalization dynamics [19]. Recent integration of Raman spectroscopy with kinetic modeling further confirmed its effectiveness in quantifying azo dye fragmentation pathways under non-thermal plasma exposure [20]. Raman spectroscopy provides direct evidence of plasma-induced structural modification of the dye molecules. The spectra show pronounced changes in the vibrational regions associated with aromatic conjugation and functional group chemistry as plasma exposure time increases. In lightly treated samples, strong and relatively sharp bands appear in the 1580–1620 cm$^{-1}$ region, characteristic of aromatic C=C stretching modes within extended π-conjugated systems, together with features in the 1200–1350 cm$^{-1}$ range attributed to C–H bending and C–C/C–O stretching vibrations. With prolonged plasma treatment, these aromatic bands progressively broaden and decrease in intensity, indicating disruption of the conjugated framework and partial loss of aromaticity. Concurrently, the 1200–1300 cm$^{-1}$ region exhibits subtle upward shifts (≈10–20 cm$^{-1}$) and relative intensity enhancement, consistent with the introduction of oxygen-containing functionalities. The emergence or strengthening of features near ~1700 cm$^{-1}$ further supports the formation of carbonyl (C=O) groups, providing clear evidence of oxidative transformation.

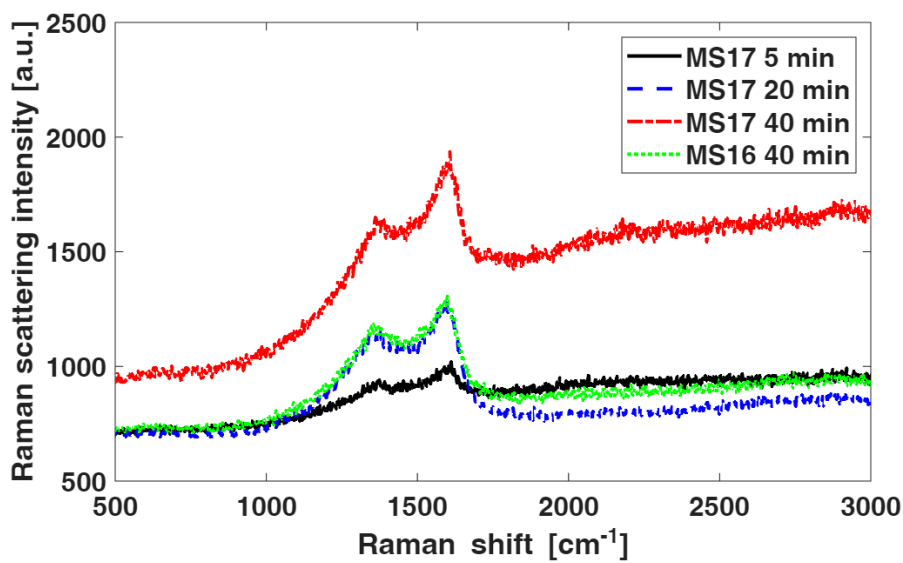

**Figure 8.** Raman spectra of MS16 and MS17 after atmospheric-pressure Ar plasma treatment for different exposure times.

These spectral modifications are consistent with the action of plasma-generated reactive oxygen and nitrogen species (RONS), including ·OH, O$_3$, and NOx, which attack the aromatic rings, induce hydroxylation and carbonyl formation, and promote fragmentation of the π-system. The progressive attenuation of the aromatic C=C stretching band reflects shortening or cleavage of conjugated segments, in agreement with the observed blue shift in fluorescence emission and the decrease in UV–Vis absorbance associated with chromophore degradation. Moreover, the concurrent solution acidification supports the formation of oxidized nitrogen species (e.g., $HNO_2$, $HNO_3$), reinforcing the oxidative environment inferred from the Raman data. Overall, vibrational evolution confirms that atmospheric-pressure plasma treatment drives stepwise oxidation and structural reorganization of the dye molecules prior to extensive fragmentation. The observed spectral evolution aligns with previously reported Raman-based mechanistic analyses of plasma-driven organic degradation, where progressive attenuation of aromatic C=C bands and emergence of carbonyl-related features signal oxidative fragmentation and partial mineralization [19,20].

## 7. Plasma–Liquid Reaction Mechanism and Coupled Kinetic–Transport Model

The atmospheric pressure argon plasma jet generates a non-thermal electron population capable of initiating electron-impact dissociation and excitation reactions. The dominant plasma-phase reactions responsible for reactive oxygen species (ROS) formation are:

$$e + Ar \rightarrow Ar^* + e \quad (1)$$
$$e + O_2 \rightarrow 2O + e \quad (2)$$
$$e + H_2O \rightarrow {}^\bullet OH + H + e \quad (3)$$
$$Ar^* + H_2O \rightarrow {}^\bullet OH + H + Ar \quad (4)$$

These reactions produce short-lived radicals (•OH, O) and longer-lived oxidants that subsequently interact with the liquid phase. The radical production rate is governed by electron density ($n_e$), electron temperature ($T_e$), and local gas composition. Reactive species generated in the plasma cross the gas–liquid interface through a combination of drift-driven penetration, diffusion, and convection induced by jet impingement. The interfacial flux $J_i$ of species i is described by:

$$J_i = k_L (C_i^* - C_i) \quad (5)$$

where $k_L$ is the liquid-side mass transfer coefficient, $C_i^*$ is the interfacial equilibrium concentration, and $C_i$ is the bulk concentration. Representative aqueous-phase reactions include:

$$O + H_2O \rightarrow 2{}^\bullet OH \quad (6)$$
$${}^\bullet OH + {}^\bullet OH \rightarrow H_2O_2 \quad (7)$$
$$2NO_2 + H_2O \rightarrow HNO_2 + HNO_3 \quad (8)$$

Plasma jet momentum induces localized convection and hydrodynamic perturbations, enhancing radical penetration depth and mixing within the near-surface liquid layer.

Within the liquid bulk, dye degradation is primarily driven by hydroxyl radicals:
$${}^\bullet OH + Dye \rightarrow Dye^\bullet + H_2O \quad (9)$$
Secondary radical chemistry includes:
$$O_2 + e_{aq}^- \rightarrow O_2^{\bullet -} \quad (10)$$

$O_2^{\bullet-} + H^+ \leftrightarrow HO_2^{\bullet}$ (11)

The equilibrium between $O_2^{\bullet-}$ and $HO_2^{\bullet}$ is pH-dependent and dynamically coupled to plasma-induced acidification.

The spatiotemporal evolution of dye concentration C(x, t) is governed by:

$\partial C/\partial t = D\nabla^2 C - u\, \partial C/\partial x - k_{OH}[\bullet OH]C - k_{O3}[O_3]C - k_{H2O2}[H_2O_2]C$ (12)

where D is the diffusion coefficient, u is the convection velocity induced by plasma jet impingement, and $k_{OH}$, $k_{O3}$, and $k_{H2O2}$ are second-order rate constants. Here, the model accounts for molecular diffusion, plasma-induced convection, nonlinear homogeneous reactions, radical recombination kinetics, and interfacial plasma-to-liquid reactive species fluxes.

The hydroxyl radical balance, ozone transport, hydrogen peroxide balance, and proton transport are given as:

$\partial[\bullet OH]/\partial t = D\nabla^2[\bullet OH] - u\, \partial[\bullet OH]/\partial x - k_{OH}[\bullet OH]C - k_{rec}[\bullet OH]^2$ (13)

$\partial[O_3]/\partial t = D\nabla^2[O_3] - u\, \partial[O_3]/\partial x - k_{O3}[O_3]C$ (14)

$\partial[H_2O_2]/\partial t = D\nabla^2[H_2O_2] - u\, \partial[H_2O_2]/\partial x + (1/2)\, k_{rec}[\bullet OH]^2 - k_{H2O2}[H_2O_2]C$ (15)

$\partial[H^+]/\partial t = D\nabla^2[H^+] - u\, \partial[H^+]/\partial x$ (16)

The boundary conditions at the plasma–liquid interface (x = 0):

$-D\, \partial[\bullet OH]/\partial x = J_{OH}$, $-D\, \partial[O3]/\partial x = J_{O3}$, $-D\, \partial[H^+]/\partial x = J_H$, $\partial C/\partial x = 0$, and

$\partial[H_2O_2]/\partial x = 0$. At the bottom boundary (x = L): $\partial[\varphi]/\partial x = 0$ (for all species), where $\varphi \in \{C, [\bullet OH], [O_3], [H_2O_2], [H^+]\}$. Initial Conditions at time t=0 are $C(x,0) = C_0$, $OH(x,0) = 0$, $O3(x,0) = 0$, $H2O2(x,0) = 0$, and $H^+(x,0) = H_0$. Kinetic and Transport Parameters are given in table (2).

Table (2). Kinetic and Transport Parameters Used in the Reaction–Diffusion Model (25 °C, Aqueous Phase).

| Parameter | Value (with unit) | Reference |
|---|---|---|
| $k_{OH}$ | $1.0 \times 10^9$ M$^{-1}$ s$^{-1}$ | [21] |
| $k_{O3}$ | $5.0 \times 10^5$ M$^{-1}$ s$^{-1}$ | [22] |
| $k_{H2O2}$ | $1.0 \times 10^{-2}$ M$^{-1}$ s$^{-1}$ | [23] |
| $k_{rec}$ | $5.5 \times 10^9$ M$^{-1}$ s$^{-1}$ | [21] |
| D (Dye) | $5.0 \times 10^{-10}$ m$^2$ s$^{-1}$ | [24] |
| $D_{OH}$ | $2.8 \times 10^{-9}$ m$^2$ s$^{-1}$ | [24] |
| $D_{O3}$ | $2.0 \times 10^{-9}$ m$^2$ s$^{-1}$ | [24] |
| $D_{H2O2}$ | $1.5 \times 10^{-9}$ m$^2$ s$^{-1}$ | [24] |
| $D_{H+}$ | $9.3 \times 10^{-9}$ m$^2$ s$^{-1}$ | [24] |

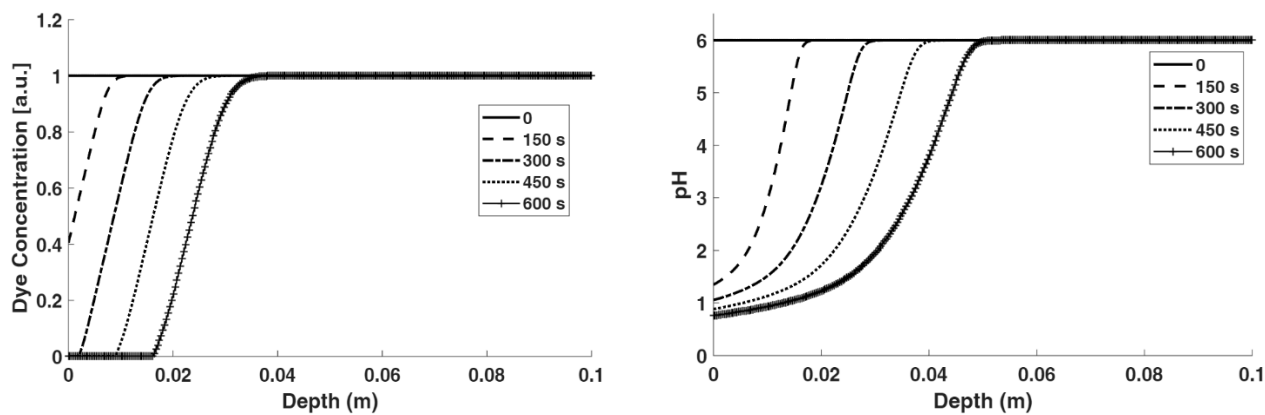

**Figure 9.** Spatiotemporal evolution of dye concentration and proton concentration (pH) predicted by the coupled reaction–diffusion–convection model.

Figure 9 presents the simulated spatiotemporal distributions of dye concentration $C(x, t)$ and proton concentration (expressed as pH) within the liquid column during plasma treatment. The calculations were performed assuming a constant interfacial hydroxyl radical flux $J_{OH} = 5 \times 10^{-7}$ mol $m^{-2}$ $s^{-1}$ and an effective convection velocity $u = 5 \times 10^{-5}$ m $s^{-1}$. The model predicts a pronounced concentration gradient originating at the plasma–liquid interface (x = 0), where reactive species are injected into the solution. Dye depletion initiates in the near-surface region and progressively propagates toward the bulk as treatment time increases. This behavior reflects the combined influence of (i) high interfacial radical density, (ii) molecular diffusion into the liquid, and (iii) plasma-induced convective transport. Simultaneously, proton concentration increases near the interface due to the dissolution of plasma-generated NOx species and secondary formation of $HNO_2$, $HNO_3$, and related acidic intermediates. The resulting acidification front follows a similar penetration profile into the liquid bulk, though with a smoother gradient owing to the higher diffusion coefficient of protons. Importantly, the simulations reveal that degradation and acidification are spatially coupled but kinetically distinct processes. Dye removal is governed primarily by localized •OH attack and oxidant transport, whereas proton accumulation reflects cumulative nitrogen oxide chemistry and proton mobility. The near-surface region therefore acts as a highly reactive boundary layer in which reaction rates exceed transport replenishment, characteristic of a reaction-dominated regime (high local Damköhler number).

Figure 10 illustrates the temporal evolution of spatially averaged dye concentration and pH under different imposed hydroxyl radical fluxes while maintaining a constant convection velocity of $u = 5 \times 10^{-5}$ m $s^{-1}$. An increase in $J_{OH}$ leads to a proportional enhancement in degradation rate, resulting in a steeper exponential decay of the normalized concentration $C/C_0$. This trend confirms that, under fixed hydrodynamic conditions, dye removal is directly controlled by the availability of interfacial reactive oxygen species. The system therefore operates in a radical-flux-controlled regime during the early treatment stage, where the apparent first-order rate constant scales with the incoming oxidant flux. In contrast, the spatially averaged proton concentration (pH evolution) exhibits only minor sensitivity to changes in $J_{OH}$. This indicates that acidification is not governed solely by hydroxyl radical chemistry but rather by nitrogen oxide dissolution pathways and secondary aqueous equilibria. The weak dependence of pH on $J_{OH}$ therefore confirms mechanistic decoupling between oxidative dye cleavage and proton accumulation. These results quantitatively demonstrate that modulation of plasma radical flux primarily tunes degradation kinetics without significantly altering the bulk acidification profile.

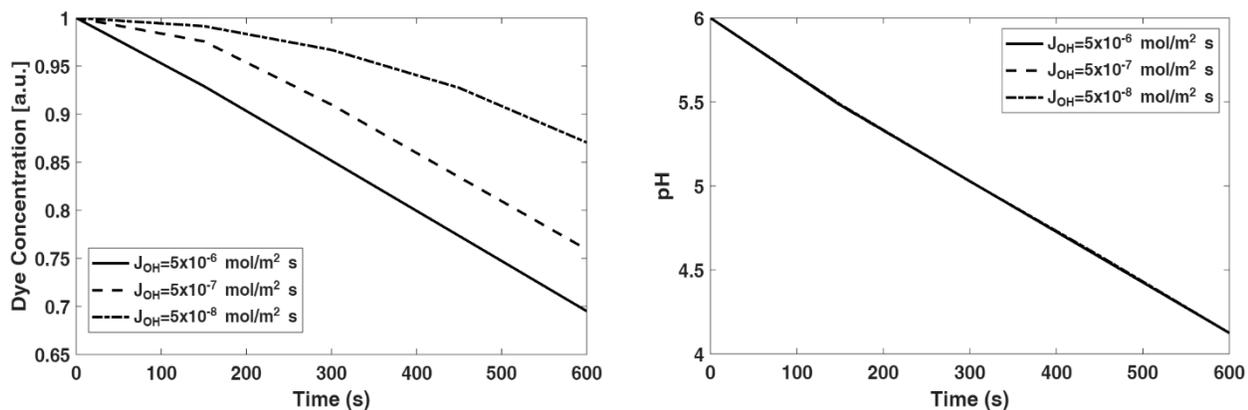

**Figure 10**. Spatially averaged dye concentration and pH evolution for varying interfacial hydroxyl radical flux $J_{OH}$ at constant convection velocity (u = 5 × 10$^{-5}$ m s$^{-1}$).

Figure 11 examines the influence of hydrodynamic transport on degradation kinetics by varying the effective convection velocity while maintaining a constant hydroxyl radical flux. At low convection velocities, degradation proceeds rapidly due to the formation of a high radical concentration boundary layer near the interface. Under these conditions, transport limitations are minimal, and the reaction rate is governed predominantly by local radical–dye interactions. This corresponds to a reaction-controlled regime characterized by a large effective Damköhler number (Da ≫ 1). As convection velocity increases, enhanced mixing distributes reactive species deeper into the bulk solution but simultaneously reduces the interfacial radical residence time. The resulting dilution of near-surface radical density decreases the local reaction rate, producing a slower overall decay of dye concentration. This shift marks a transition toward a transport-influenced regime (Da → 1), where hydrodynamic mixing competes with chemical reaction kinetics. Notably, variations in convection velocity exert a stronger influence on dye degradation than on proton accumulation, again highlighting the distinct transport properties and diffusivities of reactive oxidants and protons.

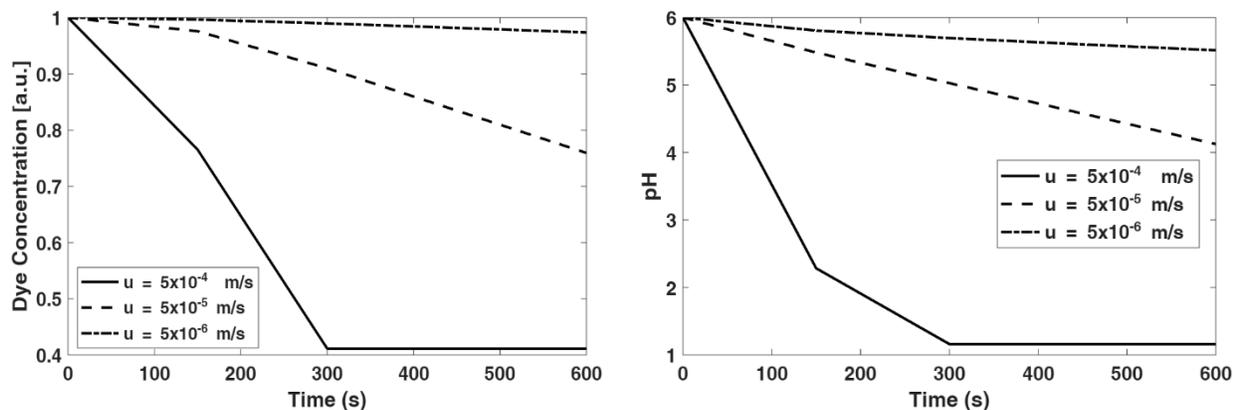

**Figure 11.** Spatially averaged dye concentration and pH evolution for varying convection velocities (u) at constant hydroxyl radical flux ($J_{OH} = 5 \times 10^{-7}$ mol m$^{-2}$ s$^{-1}$).

The theoretical results presented in Figures 9–11 provide a mechanistic basis for the experimentally observed biphasic degradation behavior. During the initial stage, high interfacial radical flux produces a reaction-controlled exponential decay: $dC/dt = -k_1 C$ and $C = C_0 \exp(-k_1 t)$, where $k_1 \propto J_{OH}$. At longer times, radical depletion, recombination losses, and transport limitations reduce the effective oxidant availability, yielding a smaller apparent rate constant: $dC/dt = -k_2 C$, where ($k_2 < k_1$). This transition reflects a dynamic shift from radical-flux-controlled oxidation to transport-influenced degradation. The model therefore reconciles experimental UV–Vis kinetics with plasma–liquid interfacial transport physics, providing a quantitative framework for scaling atmospheric-pressure plasma reactors. In future work, the present reaction–diffusion framework should be self-consistently coupled with a detailed plasma model to enable quantitative estimation of interfacial radical fluxes and electron-driven reaction rates [25–28]. Furthermore, extension to a two-dimensional formulation incorporating the Navier–Stokes equations is necessary to account for hydrodynamic perturbations, turbulence, dusts, and bubble-induced mixing effects, which may significantly influence mass transport and degradation dynamics within the plasma–liquid system [29, 35].

**Conclusion**

An atmospheric-pressure Ar plasma jet operated in a plasma–liquid discharge configuration with an immersed counter-electrode enabled efficient oxidative degradation of two azo dyes (MS16, MS17) and allowed mechanistic resolution of coupled chemistry–transport effects. Plasma treatment generated a RONS-rich environment that drove pronounced acidification, with proton concentration increasing by up to ~49 fold, consistent with NOx dissolution and secondary aqueous acid formation. UV–Vis monitoring confirmed rapid chromophore destruction and deep molecular transformation, achieving 88% (MS16) and 94% (MS17) degradation after 40 min. The degradation profiles were distinctly biphasic, supporting a transition from an initial radical-flux-controlled stage to a transport-influenced stage governed by oxidant delivery, recombination losses, and hydrodynamic mixing. Fluorescence evolution (transient enhancement and blue shift) indicated formation of oxidized emissive intermediates followed by their subsequent consumption at extended exposure. Raman spectroscopy provided molecular-level confirmation of progressive disruption of aromatic conjugation and development of oxygenated functionalities, including carbonyl formation, consistent with stepwise oxidative fragmentation of the π-system rather than simple decolorization. Differences between MS16 and MS17 establish that substituent chemistry measurably influences effective radical attack and intermediate dynamics under identical plasma conditions. Collectively, these results connect macroscopic removal metrics with spectroscopically supported molecular transformation pathways and motivate reactor/operating-condition optimization strategies that prioritize controlled oxidant flux and interfacial mass transfer for scalable plasma-based wastewater treatment.